\pdfoutput=1

\documentclass[11pt,twoside,a4paper,cmspaper,final,collab]{cms-tdr}
\def\svnVersion{8f08b24-D}\def\svnDate{2020/07/17}\def\cmsCernNoTag{CERN-EP-2020-003}\def\cmsCernDate{\today}\def\cmsMessage{"Published in Physics Letters B as \href{http://dx.doi.org/10.1016/j.physletb.2020.135578}{\doi{10.1016/j.physletb.2020.135578}.}"}

\begin{document}\cmsNoteHeader{BPH-18-002}

\hyphenation{had-ron-i-za-tion}
\hyphenation{cal-or-i-me-ter}
\hyphenation{de-vices}
\RCS$HeadURL$
\RCS$Id$

\newlength\cmsFigWidth
\newlength\cmsTabSkip\setlength{\cmsTabSkip}{1ex}
\ifthenelse{\boolean{cms@external}}{\setlength\cmsFigWidth{0.49\textwidth}}{\setlength\cmsFigWidth{0.65\textwidth}} 
\ifthenelse{\boolean{cms@external}}{\providecommand{\cmsLeft}{upper\xspace}}{\providecommand{\cmsLeft}{left\xspace}}
\ifthenelse{\boolean{cms@external}}{\providecommand{\cmsRight}{lower\xspace}}{\providecommand{\cmsRight}{right\xspace}}

\newcommand{\mmmmm}{\ensuremath{m_{\mathrm{4}\Pgm}}\xspace}
\newcommand{\mYY}{\ensuremath{m_{\PgUa\PgUa}}\xspace}
\newcommand{\monetwo}{\ensuremath{m_\text{12}}\xspace}
\newcommand{\mthreefour}{\ensuremath{m_\text{34}}\xspace}
\newcommand{\mtilde}{\ensuremath{\widetilde{m}_{\mathrm{4}\Pgm}}\xspace}
\newcommand{\deltay}{\ensuremath{\abs{\Delta y(\PgUa,\PgUa)}}\xspace}
\newcommand{\fdps}{\ensuremath{f_\text{DPS}}\xspace}
\newcommand{\sfid}{\ensuremath{\sigma_\text{fid}}\xspace}
\newcommand{\sfidSPS}{\ensuremath{\sigma_\text{fid}^\text{SPS}}\xspace}
\newcommand{\sfidDPS}{\ensuremath{\sigma_\text{fid}^\text{DPS}}\xspace}
\newcommand{\Ncorr}{\ensuremath{N^{\text{corr}}}\xspace}
\newcommand{\Nobs}{\ensuremath{N^{\text{obs}}}\xspace}
\newcommand{\ereco}{\ensuremath{\epsilon^\text{reco}}\xspace}
\newcommand{\evtx}{\ensuremath{\epsilon^\text{vtx}}\xspace}
\newcommand{\eevt}{\ensuremath{\epsilon^\text{evt}}\xspace}

\cmsNoteHeader{BPH-18-002} 
\title{Measurement of the $\PgUa$ pair production cross section and search for resonances decaying to $\PgUa\Pgm^+\Pgm^-$ in proton-proton collisions at $\sqrt{s}=13\TeV$}

\date{\today}

\abstract{
The fiducial cross section for \PgUa pair production in proton-proton collisions at a center-of-mass energy of 13\TeV in the region where both \PgUa mesons have an absolute rapidity below 2.0 is measured to be $79 \pm 11\stat \pm 6\syst\pm 3~(\mathcal{B})$\unit{pb} assuming the mesons are produced unpolarized. The last uncertainty corresponds to the uncertainty in the \PgUa meson dimuon branching fraction. The measurement is performed in the final state with four muons using proton-proton collision data collected in 2016 by the CMS experiment at the LHC, corresponding to an integrated luminosity of 35.9\fbinv. This process serves as a standard model reference in a search for narrow resonances decaying to $\PgUa\Pgm^+\Pgm^-$ in the same final state. Such a resonance could indicate the existence of a tetraquark that is a bound state of two \cPqb quarks and two \cPaqb antiquarks. The tetraquark search is performed for masses in the vicinity of four times the bottom quark mass, between 17.5 and 19.5\GeV, while a generic search for other resonances is performed for masses between 16.5 and 27\GeV. No significant excess of events compatible with a narrow resonance is observed in the data. Limits on the production cross section times branching fraction to four muons via an intermediate \PgUa resonance are set as a function of the resonance mass. 
}

\hypersetup{
pdfauthor={CMS Collaboration},%
pdftitle={Measurement of the Y(1S) pair production cross section and search for resonances decaying to Y(1S) mu+ mu- in proton-proton collisions at sqrt(s) = 13 TeV},
pdfsubject={CMS},
pdfkeywords={CMS, Upsilon}}

\maketitle 

\section{Introduction}
\label{sec:intro}

Quarkonium pair production is an important probe of both perturbative and nonperturbative processes in quantum chromodynamics.  
Experimental studies of this process can provide valuable information about the underlying mechanisms of particle production and improve our understanding 
of numerous physics processes that are treated as backgrounds in searches and measurements.
Quarkonium pairs may originate from single-parton scattering (SPS) or double-parton scattering (DPS). These production mechanisms can be separated experimentally 
since the DPS production is characterized, among other features, by more forward and separated mesons. 
The analysis of nonperturbative effects is easier for quarkonium states composed of \cPqb quarks, 
as their large masses allow them to be approximated as nonrelativistic systems~\cite{Lansberg:2019adr}. 
The CMS Collaboration observed for the first time the production of a pair of \PgUa mesons, using proton-proton data collected at a center-of-mass energy of 8\TeV~\cite{Khachatryan:2016ydm}. 
This Letter presents a measurement of the \PgUa pair production cross section at a center-of-mass energy of 13\TeV. 
The cross section is measured in the fiducial region where both \PgUa mesons have an absolute rapidity below 2.0, using 
the final state 
with four muons. Additionally, the DPS contribution to the process is measured for the first time.

The \PgUa pair production can serve as a reference in searches for tetraquarks or generic resonances with masses close to twice the 
\PgUa meson mass. 
A light resonance decaying to a \PgUa meson and a pair of leptons might be the signature of
a tetraquark characterized as a bound state of two \cPqb quarks and two \cPaqb antiquarks, especially if its mass is below twice the $\eta_\cPqb$ mass~\cite{Yuqi:2011gm,Berezhnoy:2011xn,Dobrescu:2014fca,Karliner:2016zzc,Chen:2016jxd,Wang:2017jtz,Wu:2016vtq,Anwar:2017toa,Hughes:2017xie,Esposito:2018cwh,Bai:2016int}.
In this Letter, in addition to the measurement of the \PgUa pair production cross section, we describe a search for tetraquarks with masses between 17.5 and 19.5\GeV, since
$\cPqb \cPqb \cPaqb \cPaqb$ tetraquarks would be expected to have a mass around four times that of the bottom quark.
A generic search for narrow resonances with mass between 16.5 and 27\GeV and decaying to a \PgUa meson and a pair of muons is also presented. The final state is the same
as for the measurement of the \PgUa pair production cross section, and a similar event selection is used.
The \PgUa pair production is a background to the resonance search. 
 
The LHCb Collaboration searched for $\cPqb\cPqb\cPaqb\cPaqb$ tetraquarks using data collected at center-of-mass energies of 7, 8, and 13\TeV, without finding any hint of a signal~\cite{Aaij:2018zrb}. 
This analysis
probes a kinematic region that is not accessible with the LHCb detector and extends the covered mass range in the context of the generic search.

The \PgUa pair production fiducial cross section measurement and the resonance search are based on proton-proton collision data collected in 2016 at a center-of-mass energy of 13\TeV by the CMS experiment at the CERN LHC,
corresponding to an integrated luminosity of 35.9\fbinv. 

\section{The CMS detector}
\label{sec:detector}

The central feature of the CMS apparatus is a superconducting solenoid of 6\unit{m} internal diameter, providing a magnetic field of 3.8\unit{T}. Within the solenoid volume, there are a silicon pixel and strip tracker, a lead tungstate crystal electromagnetic calorimeter, and a brass and scintillator hadron calorimeter, each composed of a barrel and two endcap sections. Forward calorimeters extend the pseudorapidity coverage provided by the barrel and endcap detectors. Muons are detected in gas-ionization chambers embedded in the steel flux-return yoke outside the solenoid. Events of interest are selected using a two-tiered trigger system~\cite{Khachatryan:2016bia}. A more detailed description of the CMS detector, together with a definition of the coordinate system used and the relevant kinematic variables, can be found in Ref.~\cite{Chatrchyan:2008zzk}.

Muons are measured in the range $\abs{\eta} < 2.4$, with detection planes made using three technologies: drift tubes, cathode strip chambers, and resistive plate chambers. Matching muons to tracks measured in the silicon tracker results in a relative \pt resolution in the range 0.8--3.0\% for muons with \pt less than 10\GeV~\cite{Sirunyan:2018fpa}.

\section{Simulated samples}
\label{sec:simulation}

The \PgUa pair production signal is simulated using the \PYTHIA 8.226 generator~\cite{Sjostrand:2014zea}, separately for the 
SPS and DPS mechanisms, under the assumption that the mesons are produced unpolarized. 
The DPS sample is produced by generating two hard interactions with color-singlet production of bottomonium states via 
$\cPg\cPg \to \cPqb\cPaqb$ or color-octet production of bottomonium states via $\cPq\cPaq \to \cPqb\cPaqb$. 
The invariant mass distribution of the meson pair and of the rapidity separation between the mesons are used to extract the fraction of DPS production, as detailed in 
Section~\ref{sec:xs}. For this measurement, the distributions of these variables for the SPS process are taken from the next-to-leading-order (NLO*) calculation with a cutoff  
color-singlet mechanism (CSM)~\cite{Baier:1981uk,Lansberg:2013qka,Lansberg:2014swa} using \textsc{HELAC-Onia} 2.0.1~\cite{Shao:2012iz,Shao:2015vga}.

The signal of a narrow resonance decaying to a \PgUa meson and a pair of muons is modeled using different physics assumptions depending on the nature of the resonance:  
\begin{itemize}
\item a bottomonium state with the properties of the \Pbgci, assuming a phase-space decay to a \PgUa meson and a pair of muons, using the \PYTHIA 8.226 generator;
\item a scalar particle produced in gluon fusion, using the \textsc{JHUGen} generator~\cite{Gao:2010qx,Bolognesi:2012mm,Anderson:2013afp,Gritsan:2016hjl};
\item a pseudoscalar particle produced in gluon fusion, using the \textsc{JHUGen} generator;
\item a spin-2 particle produced in gluon fusion, using the \textsc{JHUGen} generator.
\end{itemize}
The signals are generated assuming the narrow-width approximation. 
The \Pbgci sample is used to model the tetraquark signal, for which no dedicated generator exists. The other samples
correspond to the signals in the generic search over an extended mass range.
For each model, four resonance mass values are simulated: 14, 18, 22, and 26\GeV. Since the signal acceptance falls steeply 
 around and below 14\GeV in the
simulated samples, the probed mass range in this analysis is restricted to stay well above this mass threshold. The different mass points are used to interpolate and extrapolate the signal 
model over the whole mass range. 

The \PYTHIA generator with the tune CUETP8M1~\cite{Khachatryan:2015pea} is used to model the parton shower and hadronization processes. Generated events are processed through a simulation of the CMS detector based on \GEANTfour~\cite{Agostinelli:2002hh}.

\section{Event selection criteria}
\label{sec:selection}

The event reconstruction is based on the particle-flow algorithm~\cite{Sirunyan:2017ulk}, which identifies individual particle candidates using information 
from all the individual subdetectors.
Muons are reconstructed by combining information from the silicon tracker and the muon system~\cite{Sirunyan:2018fpa}. 

Events are selected with a trigger that requires the presence of three muons. 
Among these muons, two must have an invariant mass compatible with a \PgU resonance ($8.5 < m_{2\Pgm} < 11.4 \GeV$) 
at trigger level, and the dimuon vertex fit probability, calculated using the $\chi^2$ and the number of degrees of freedom of the fit, must be  
greater than 0.5\%.  

Offline, we require each event to have four reconstructed muons with $\pt>2\GeV$ and $\abs{\eta}<2.4$. 
These muons are required to satisfy the
global or particle-flow muon identification criteria described in Ref.~\cite{Sirunyan:2018fpa}.
About 25\% of simulated signal events and about 30\% of data events have
more than four such muons.
Possible combinations of four muon tracks are refit with a constraint to
come from a common vertex, and the $\chi^2$ probability of the fit is
determined.
The combination of four muons with the largest $\chi^2$ probability is chosen. 
For simulated signal events with more than four reconstructed muons, the correct muons are chosen in about 98\% of cases. 
Among the four muons, at least three need to be associated with the trigger-level objects.  
At least two muons must be associated with the objects that passed the \PgU mass compatibility and vertex criteria of the trigger, 
and they are paired together. 
If there are more than two such muons, which happens for 2 to 35\% of simulated signal events  
depending on the resonance mass,
those that have opposite-sign (OS) charges and an invariant mass closest to the world-average \PgUa mass~\cite{PhysRevD.98.030001} are paired together.

After selecting the best combination of four muons with $\pt>2\GeV$, the \pt threshold is raised to 2.5\GeV for the selected muons. 
The final selection requiring $\pt>2.5\GeV$ 
reduces the background from misidentified muons by about a factor of two. 
The muons are required to satisfy the 
medium muon identification criteria described in Ref.~\cite{Sirunyan:2018fpa}. 
Both pairs of muons have to be composed of OS muons. The vertex fit $\chi^2$ probability of the four muons is 
required to be greater than 5\%, 
whereas that of the \PgUa candidate is required to be above 0.5\%, similar to the requirement already imposed at trigger level. 
The muons are required to be separated from each other by at least $\Delta R = \sqrt{\smash[b]{(\Delta \eta)^2 + (\Delta \phi)^2}} = 0.02$, where $\Delta \eta$ and $\Delta \phi$ are the differences in pseudorapidity and azimuthal angle between the muons.  
The positively (negatively) charged muon from one of the pairs can be paired with the negatively (positively) charged 
muon of the other pair to form so-called alternative pairs of OS muons. If one of these alternative pairs has an invariant mass compatible with a \JPsi particle 
within two standard deviations of the experimental resolution, which ranges between about 0.03 and 0.12\GeV depending on the muon pair kinematics, the event is discarded from the analysis.
Events are also discarded if they contain two OS pairs of muons with invariant mass less than 4\GeV.

The selection criteria detailed above are common for the measurement of the \PgUa pair production cross section and the search for a resonant signal.
The criteria that differ between the measurement and the search are described in the following. In the measurement of the \PgUa pair fiducial cross section, 
the reconstructed absolute rapidity of both muon pairs is required 
to be less than 2.0. In addition, for muons with $\abs{\eta}<0.9$, the \pt threshold is raised to 3.5\GeV.
Central muons with transverse momentum below 3.5\GeV have a high probability of being absorbed in the calorimeter or  
undergoing significant 
multiple scattering before reaching 
the muon detectors. This selection criterion reduces the systematic uncertainty in the muon reconstruction 
related to the detector simulation. It is, however, not used in the resonant search 
because it would strongly reduce the signal acceptance for the 
lower-mass signal range. 
In the resonance search, the invariant mass of the \PgUa candidate is required 
to be within two standard deviations of the experimental resolution from the \PgUa mass~\cite{PhysRevD.98.030001}, 
where the resolution varies between about 0.06 and 0.15\GeV depending on the event. 

The mass range of interest is known a priori for the search of a $\cPqb \cPqb \cPaqb \cPaqb$ tetraquark signal.
In this case,  
all the selection 
criteria described above have been determined and fixed in a blinded way, 
using simulation and without looking at data events with four muons having an invariant mass between 17.5 and 19.5\GeV.

\boldmath
\section{Measurement of the \texorpdfstring{$\PgUa$}{Upsilon(1S)} pair production cross section}
\label{sec:xs}
\unboldmath

The methodology used to measure the \PgUa pair production cross section is detailed in Section~\ref{sec:methodologyxs}. 
After discussing the systematic uncertainties in Section~\ref{sec:systematics1}, the results of the measurement of the inclusive \PgUa
pair production fiducial cross section are presented
in Section~\ref{sec:results1}. Nonisotropic decays of the \PgUa mesons would
change the measured cross section. Section~\ref{sec:pola} describes how the cross section would vary for nonzero values of the polarization parameters.
Finally, the DPS and SPS mechanisms can be separated experimentally by measuring the \PgUa pair production cross section
in bins of the rapidity difference between the mesons, \deltay, and of the invariant mass of the meson pairs, \mYY. A measurement of the DPS-to-inclusive cross section ratio in the fiducial region
is presented in Section~\ref{sec:fdps}.

\subsection{Methodology}\label{sec:methodologyxs}

The \PgUa pair production cross section is measured in the fiducial region where both mesons have an 
absolute rapidity below 2.0. No other requirement is applied to define the fiducial region.   
The fiducial cross section, \sfid, can be expressed as:
\begin{equation}\label{eq:Ncorr}
\sfid=\frac{\Ncorr}{\lumi\mathcal{B}^2},
\end{equation} 
where $\Ncorr$ is the number of signal events corrected for the acceptance and efficiency of the selection, 
$\lumi$ is the integrated luminosity, and $\mathcal{B}$ stands for $\mathcal{B}(\PgUa\to\Pgm^+\Pgm^-)=(2.48\pm 0.05)\%$~\cite{PhysRevD.98.030001}, which is the branching fraction of 
the \PgUa meson decay to a pair of muons. 
To extract \Ncorr from the data, we perform an extended unbinned two-dimensional (2D) maximum likelihood fit of the invariant mass
distributions of two OS muon pairs, where all events are weighted for the acceptance and efficiency on an event-by-event basis by the weight $\omega$, defined as:
\begin{linenomath}
\begin{equation}
\label{eq:corr}
\begin{split}
& \omega= \big[A_1 A_2 \ereco_1 \ereco_2 \big(1-(1-\evtx_1)(1-\evtx_2)\big)\eevt \big]^{-1}, 
\end{split}
\end{equation}
\end{linenomath}
where the different terms are described below:
\begin{itemize}
\item $A$, the probability for a \PgUa meson with an absolute rapidity below 2.0 and decaying to a pair of 
muons to have two muons in the geometrical acceptance of the detector (muon $\abs{\eta}<2.4$); No strong correlation between
the acceptance values of the two mesons are found with a closure test
described in Section~\ref{sec:systematics1}, and the total acceptance is therefore computed
as the product of the per-meson weights;
\item \ereco, the probability for a \PgUa meson with an absolute rapidity below 2.0 and decaying to a pair of
muons each with $\abs{\eta}<2.4$ to have two reconstructed muons passing the identification and kinematic criteria listed in Section~\ref{sec:selection};
\item \evtx, the probability for a \PgUa meson passing the acceptance reconstruction criteria outlined in items 2 and 3 to have a vertex fit $\chi^2$ probability above 0.5\%;
\item \eevt, the probability for an event where both \PgUa candidates pass all the criteria of items 2 and 3, 
and at least one of them passes the vertex fit $\chi^2$ probability criterion of item 4, to pass the following event-level criteria: the trigger requirements, the four-muon vertex fit $\chi^2$ probability above 5\%, and the absence of OS dimuon pairs with an invariant mass within two standard deviations of the world-average \JPsi meson mass~\cite{PhysRevD.98.030001}. 
\end{itemize}
The first three items in the above list are calculated as a function of the \PgUa rapidity and \pt. The values of $A$, \ereco, and 
\evtx, range 
between 0.47 and 1.00, 0.23 and 0.88, and 0.81 and 0.98, respectively, depending on the \PgUa rapidity and \pt. 
The factor \eevt is calculated as a function of the \pt of both \PgUa candidates, and ranges between 0.33 and 0.65.
The subscript indices in Eq.~(\ref{eq:corr}) indicate the \PgUa candidate to which the weight corresponds. The factor \evtx enters the 
formula differently from the other acceptance and efficiency terms  
because the dimuon vertex fit $\chi^2$ probability criterion needs 
to be satisfied by at least one of the two \PgUa candidates, but not necessarily by both. 
The weight $\omega$ is computed on an event-by-event basis, using the kinematic quantities of the reconstructed \PgUa candidates in data. They are estimated 
from simulation as efficiency maps and are similar for the SPS and DPS production modes, despite different correlations between the mesons. Data-to-simulation corrections 
for the trigger and muon identification efficiencies are taken into account in the computation of \Ncorr. 

In about 3\% of cases, the four reconstructed muons are not correctly paired in the SPS and DPS \PgUa pair simulations.
These events cannot
be identified as part of the signal by the 2D fit since their distribution is similar to that of the floating combinatorial background. Therefore,
the value \Ncorr extracted from the fit
is corrected by $+3\%$ to take into account these mispairings.

In the 2D fit, the muons are paired as described in Section \ref{sec:selection}, and the invariant 
masses of the two pairs are randomly denoted \monetwo and \mthreefour. 
The signal model corresponds to $\PgUa+\PgUa$ events, whereas the background model is the sum of the following physics processes:
\begin{itemize}
\item $\PgUb+\PgUb$;
\item $\PgUc+\PgUc$;
\item $\PgUb+\PgUa$;
\item $\PgUc+\PgUa$;
\item $\PgUa+\textrm{combinatorial background}$;
\item $\PgUb+\textrm{combinatorial background}$;
\item $\PgUc+\textrm{combinatorial background}$;
\item combinatorial background + combinatorial background.
\end{itemize}

\begin{figure}[hbpt]
\centering
        \includegraphics[width=0.49\textwidth]{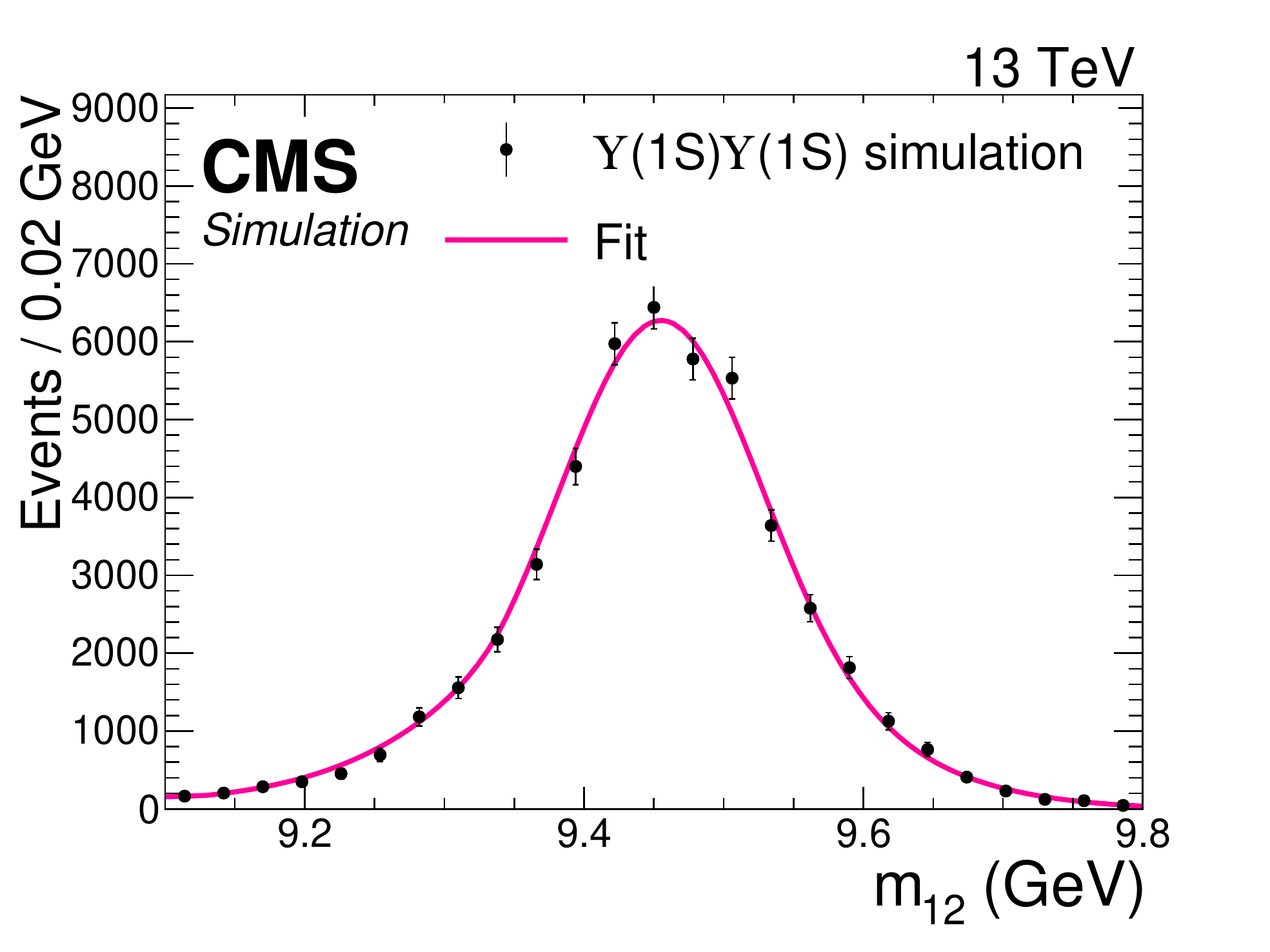}
    \caption{Projection of the 2D fit (line) to the \monetwo invariant mass distribution (points) for the SPS $\PgUa\PgUa$ simulation. The vertical bars on the points show the statistical uncertainty only. The mass distribution is modeled with the sum of two Crystal Ball functions with the same mean. }
    \label{fig:2DYY}
\end{figure}

The shape of the invariant mass distribution for the \PgUa component is determined from a 2D fit of the 
two dimuon invariant masses in the \PgUa pair SPS simulation. The results are verified to be compatible with those of a fit performed 
using the simulated DPS events, even if the muon rapidity distributions differ between production modes. 
The \monetwo and \mthreefour distributions are fitted with the sum of two same-mean Crystal Ball functions, which correspond to a power law tail added to a Gaussian core. 
This allows the radiative 
tails of the distributions to be well modeled.  
Figure~\ref{fig:2DYY} shows the projection of the 2D fit on the \monetwo axis for $\PgUa\PgUa$ simulated events. The projection on the \mthreefour axis is statistically 
identical and therefore not shown.
The fitted mean of the Crystal Ball functions in simulation is compatible within one standard deviation with the world-average 
mass of the \PgUa meson, while the full width at half maximum is about 0.19\GeV, which is several orders of 
magnitude larger than the world-average width of the \PgUa meson~\cite{PhysRevD.98.030001} because of the limited 
detector resolution.

The contributions from \PgUb and \PgUc mesons are small, and the dimuon invariant mass distributions for these mesons are taken
from a control region in data with events with two muons and two additional tracks that
do not correspond to muon candidates.
Both processes are modeled with a Gaussian function.

The combinatorial background components in the \monetwo and \mthreefour distributions are modeled with second-order Chebychev polynomials with identical parameters. 
The number of degrees of freedom has been determined with a Fisher F-test~\cite{Fisher}, where the distribution of the combinatorial background is found by inverting the muon 
pair association in the signal region. 
The parameters of the polynomial are free to float in the 2D fit to data in the signal region, detailed in Section~\ref{sec:results1}.

In the 2D fit to the data performed in the signal region, the free parameters are the normalizations of all the processes 
and the parameters of the combinatorial background mass distribution. The function parameters of the \PgUa, \PgUb, and \PgUc 
signal shapes  
are constrained within their uncertainties.

\subsection{Systematic uncertainties}
\label{sec:systematics1}

The normalization uncertainties that affect the measurement are the following:
\begin{itemize}
\item 2.5\% uncertainty in the integrated luminosity for the 2016 running period~\cite{CMS-PAS-LUM-17-001}, which appears in Eq.~(\ref{eq:Ncorr}).
\item 0.5\% uncertainty per muon in the efficiency of the muon identification and tracking, measured with a tag-and-probe method~\cite{Sirunyan:2018fpa}. It sums up to 2\% per event because the uncertainties are assumed to be correlated for the four muons since they mostly originate from the same source. This uncertainty is related to the term \ereco in the weight $\omega$.
\item 1\% uncertainty in the vertex fit $\chi^2$ probability criterion, determined by comparing background-subtracted observed and simulated distributions of the vertex fit $\chi^2$ probability for events with a \PgUa
meson and two nearby tracks. This uncertainty is related to the term \evtx in the weight $\omega$.
\item 2\% uncertainty per muon matched to trigger objects in the trigger efficiency, measured with a tag-and probe method, summing up to 6\% per event because the uncertainties are assumed to be correlated for the three muons required at trigger level. This uncertainty is related to the term \eevt in the weight $\omega$. 
\end{itemize}
These normalization uncertainties propagate directly into identical uncertainties in
the \PgUa pair production cross section.  Additionally, the uncertainty of 2\% in the $\mathcal{B}(\PgUa\to\Pgm^+\Pgm^-)$ branching
fraction, which is used to compute \Ncorr based on Eq.~(\ref{eq:Ncorr}), results in a 4\% uncertainty in the \PgUa pair production cross section measurement.

The parameters of the combinatorial background are freely floating, while the parameters of the
$\PgUa\PgUa$ distributions are constrained within the uncertainties obtained from the fit to simulated events. An 
uncertainty of 0.2\% in the muon momentum scale is propagated as an uncertainty in the mean of the 
\PgUa model. These uncertainties in the signal and background model together contribute an uncertainty of 1.5\% in the \PgUa pair production cross section measurement.

The consistency of the method to obtain \Ncorr is checked by applying the efficiency and acceptance weights to the events selected in simulation, and comparing the computed \Ncorr to the number of events generated in the fiducial
region before applying any selection criterion.
This test is performed for both the SPS and DPS simulations using the correction maps derived from  
one sample, the other one, or their combination.  Using the combined map, the weighted DPS yield has a deviation of $(-1.3\pm 3.7)\%$ with respect to the generated yield, and the corresponding deviation for the SPS sample
is $(-0.6\pm 1.5)\%$. The level of closure is similarly good for both production modes despite average 
event weights differing by more than a factor of 3 because 
of the kinematic differences. The weighted number of data events used to compute the \PgUa pair production cross section
is increased by 1\% to allow for a potential nonclosure, and an uncertainty of 1.5\% is
associated with this correction.

The systematic uncertainties are summarized in Table~\ref{tab:syst}.

\begin{table*}
\topcaption{Systematic uncertainties considered in the \PgUa pair production cross section measurement. The last column gives the associated absolute uncertainty in the measurement of \sfid. }
\label{tab:syst}
\centering
\begin{tabular}{lcc}
Uncertainty source & Uncertainty (\%) & Impact on \sfid (\unit{pb}) \\
\hline
Integrated luminosity & 2.5 & 2.0 \\
Muon identification & 2.0 & 1.6 \\
Trigger & 6.0 & 4.7  \\
Vertex probability & 1.0 & 0.8  \\
$\mathcal{B}(\PgUa\to\Pgm^+\Pgm^-)$ & 4.0 & 3.2 \\
Signal and background models & 1.2 & 1.0  \\
Method closure & 1.5 & 1.2  \\[\cmsTabSkip]
Total & 8.1 & 6.4 \\
\end{tabular}
\end{table*}

\subsection{Measurement of the fiducial cross section}
\label{sec:results1}

\begin{figure}[h!]
    \centering
        \includegraphics[width=0.47\textwidth]{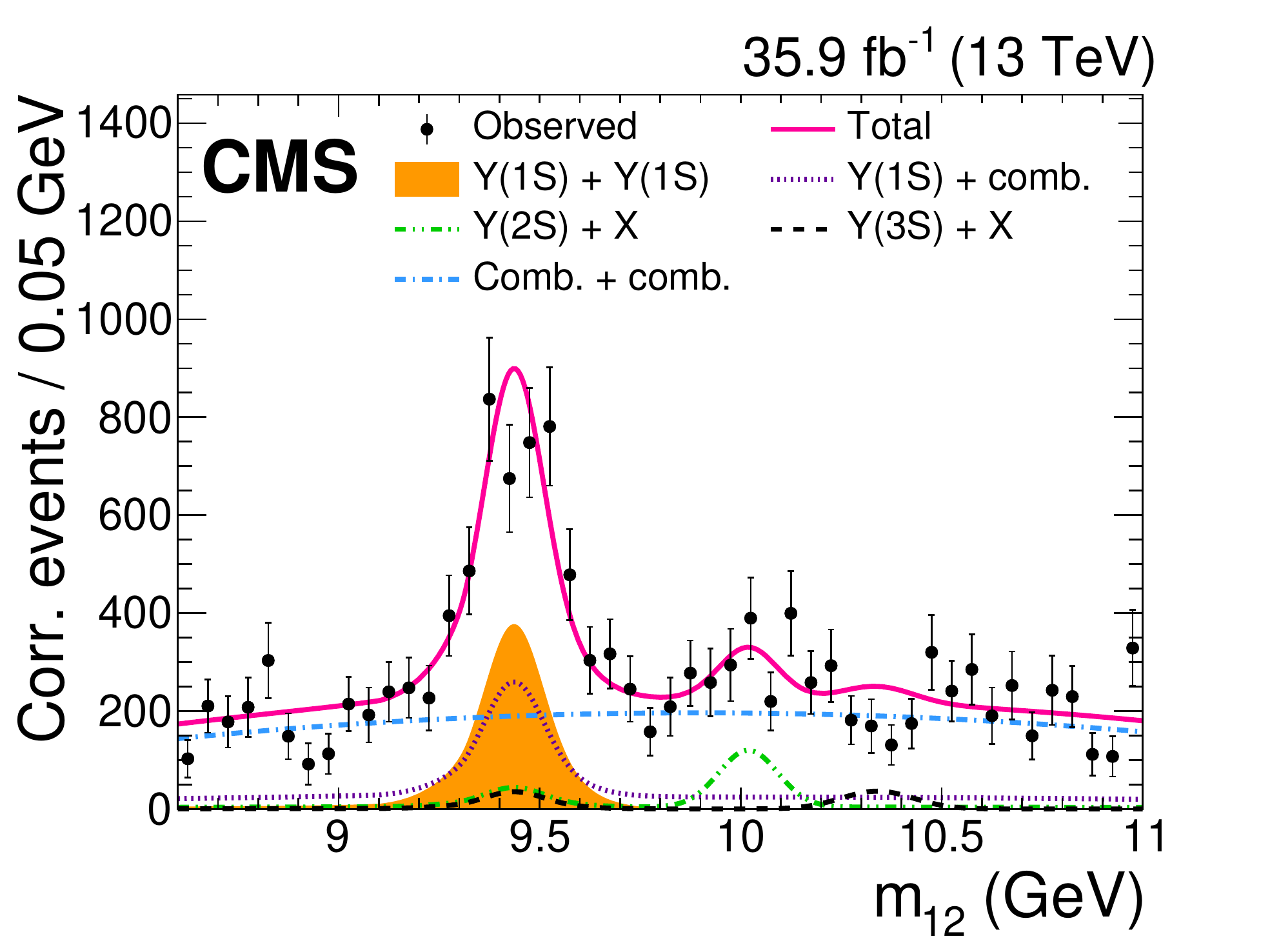}
        \includegraphics[width=0.47\textwidth]{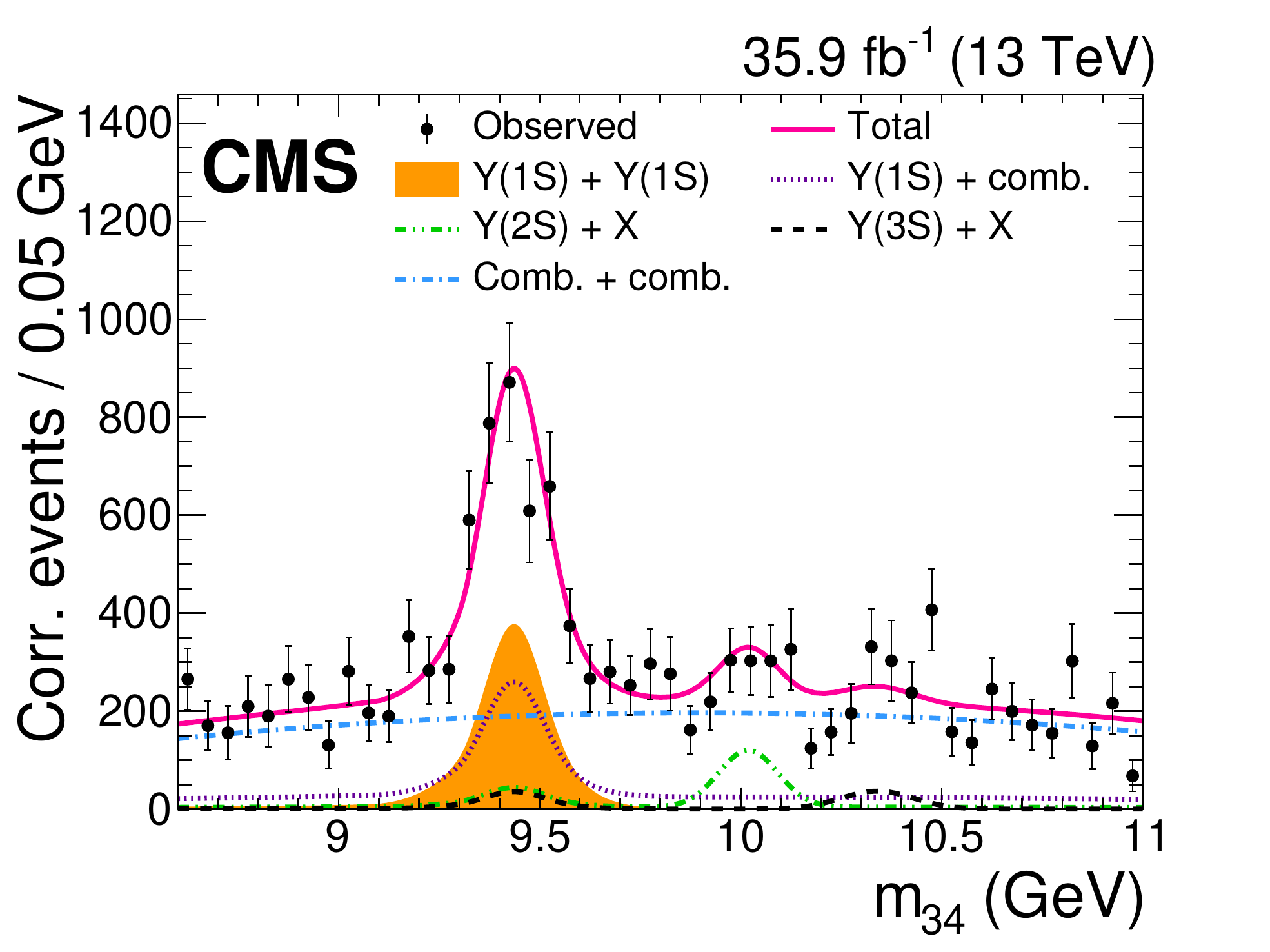}
    \caption{The two projections and the result of the 2D fit to the muon pair invariant masses. Each event is corrected for acceptance and efficiency. The \PgUa pair production signal is shown as a filled area. The contributions from the combinatorial background, and from events with a \PgUa meson and a pair of combinatorial muons, with a \PgUb meson and two reconstructed muons from any origin, and with a \PgUc meson and two reconstructed muons from any origin, are overlaid. }
    \label{fig:YY}
\end{figure}

The 2D unbinned fit to the \monetwo vs. \mthreefour distribution yields $\Ncorr=1740\pm240$ for the $\PgUa\PgUa$ process. The projections on both dimensions with all the fit components
are shown in Fig.~\ref{fig:YY}. This number of events can be translated into an inclusive cross section
for the $\PgUa\PgUa$ process in the fiducial region
defined such that both \PgUa mesons have an absolute rapidity below 2.0.
Taking into account the statistical and systematic uncertainties described in Section \ref{sec:systematics1}, and assuming unpolarized \PgUa mesons, 
the inclusive fiducial cross section is measured to be:
\begin{linenomath} 
\begin{equation}
\sfid=79 \pm 11\stat \pm 6\syst\pm3~(\mathcal{B}) \  \unit{pb},
\end{equation}
\end{linenomath}
where the last uncertainty comes from the uncertainty in the \PgUa dimuon branching fraction. 

The CMS Collaboration previously measured, in the same fiducial region, the $\PgUa\PgUa$ production cross section at a center-of-mass energy of 8\TeV to be $69\pm13\stat\pm7\syst\pm3(\mathcal{B})$\unit{pb}~\cite{Khachatryan:2016ydm}. Assuming all uncertainties are uncorrelated with those in the result presented in this Letter except that in the branching fraction of the \PgUa meson to muons,
the measured ratio of the cross section at a center-of-mass of 13\TeV to that at 8\TeV is $1.14\pm 0.32$, 
where the uncertainty includes both the statistical and systematic components. The \PYTHIA generator predicts a ratio of 2.1 for DPS production,
and 1.6 for the SPS production. Taking the fraction of the DPS mechanism in the total cross section $\fdps=(39\pm 14)\%$ at a center-of-mass energy of 13\TeV, as measured in Section~\ref{sec:fdps}, the cross section ratio predicted by \PYTHIA is
$1.79\pm0.27$. Combining the uncertainties in quadrature, the prediction is within two standard deviations of the measurement.

\begin{figure}[h!]
    \centering
        \includegraphics[width=0.47\textwidth]{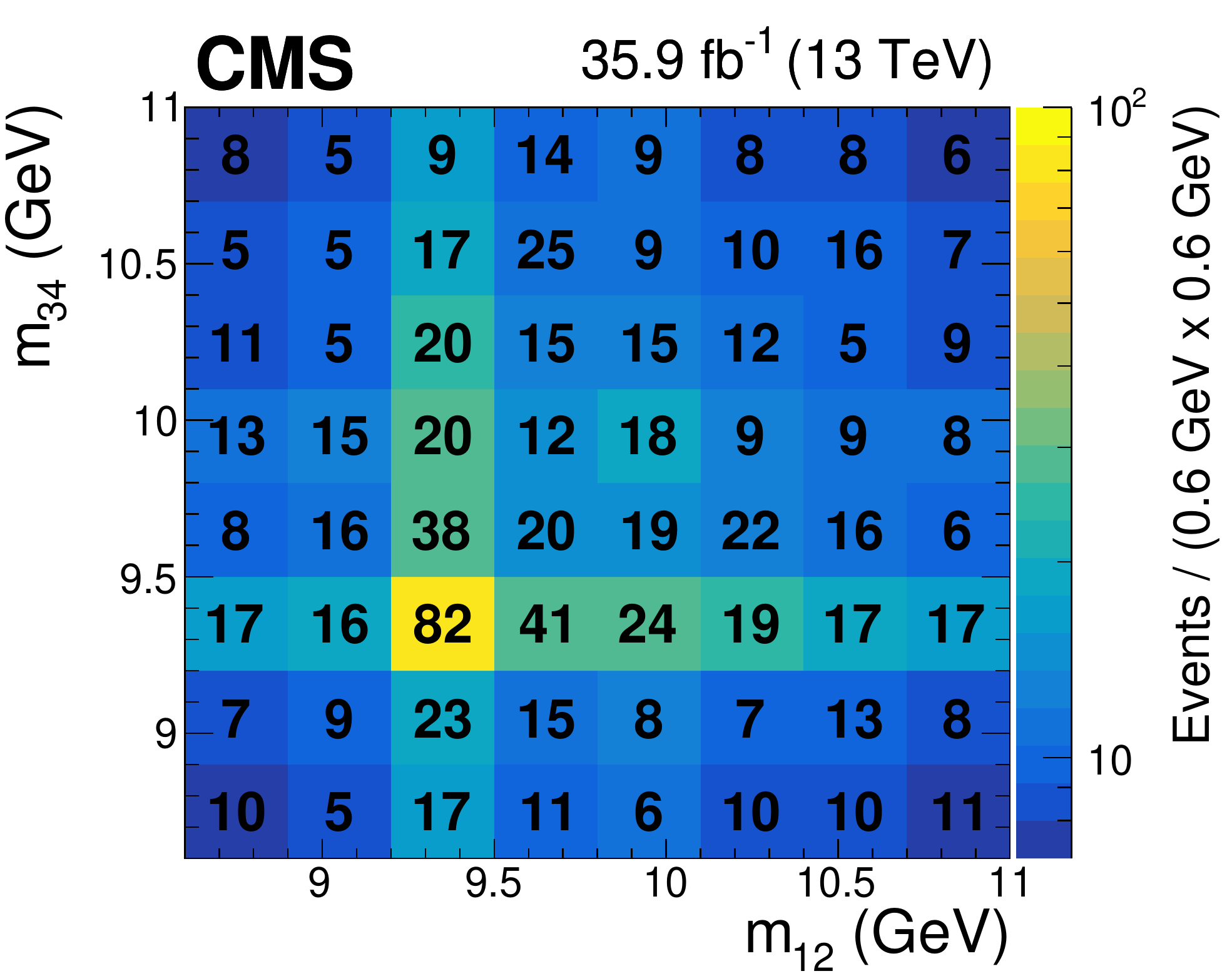}
    \caption{The number of data events in each $0.6\GeV\times 0.6\GeV$ bin of the \monetwo vs. \mthreefour distribution is shown. The results of the maximum-likelihood fit to the signal+background model are given by the colors, using the color scale to the right of the plot. }
    \label{fig:d2}
\end{figure}

Another unbinned extended maximum likelihood fit is performed to extract the number of $\PgUa\PgUa$ events observed 
in data after the selection. The $\PgUa\PgUa$ unweighted signal yield is obtained from 
a fit where all observed events have a weight of 1.0. 
For this fit, a separate signal shape is determined by fitting the \monetwo and \mthreefour distributions in the unweighted simulation. The absence of weighting does not significantly modify the signal distribution. 
The unweighted event yields are given for all processes in Table~\ref{tab:yields}. 
There is no evidence for the simultaneous production of two excited states of the \PgU meson, but excesses with a significance lower than two standard deviations indicate the possible presence of $\PgUa\PgUb$ and $\PgUa\PgUc$ events.
The number of events from data in the \monetwo vs. \mthreefour distribution is shown 
in Fig.~\ref{fig:d2}, along with the results of the fit to the signal+background model, 
using the color scale to the right of the plot. 

\begin{table}
\topcaption{The unweighted number of events for each of the processes from the fit to the \monetwo and \mthreefour distributions without acceptance nor efficiency corrections. }
\label{tab:yields}
\centering
\begin{tabular}{lc}
Process & Uncorrected yield \\
\hline
$\PgUa+\PgUa$ & $111\pm16$\\
$\PgUb+\PgUb$ & $3.6~^{+4.4}_{-3.6}$ \\
$\PgUc+\PgUc$ & $1.1~^{+1.4}_{-1.1}$ \\
$\PgUa+\text{combinatorial}$ & $166\pm33$ \\
$\PgUb+\text{combinatorial}$ & $25\pm18$ \\
$\PgUc+\text{combinatorial}$ & $1.1~^{+11}_{-1.1}$ \\
$\PgUb+\PgUa$ & $19\pm10$ \\
$\PgUc+\PgUa$ & $17\pm11$ \\
Combinatorial + combinatorial & $561\pm41$ \\
\end{tabular}
\end{table}

\subsection{Effect of the polarization}
\label{sec:pola}

The acceptance and efficiency corrections have been computed assuming negligible polarization of the \PgUa mesons.
A different assumption on the polarization can change the measured fiducial cross section. The polarization of the \PgUa states
affects the angular distributions of the leptons produced in the $\PgUa\to\Pgm^+\Pgm^-$ decays 
through the following formula~\cite{Faccioli:2010kd}:
\begin{linenomath}
\ifthenelse{\boolean{cms@external}}
{ 
\begin{multline*}
\frac{\rd^2N}{\rd\cos\theta~\rd\phi}\propto \frac{1}{3+\lambda_\theta}(1+\lambda_\theta\cos^2\theta + \lambda_\phi\sin^2\theta\cos 2\phi \\ 
+ \lambda_{\theta\phi}\sin 2\theta\cos\phi),
\end{multline*}
} 
{ 
\begin{equation*}
\frac{\rd^2N}{\rd\cos\theta~\rd\phi}\propto \frac{1}{3+\lambda_\theta}(1+\lambda_\theta\cos^2\theta + \lambda_\phi\sin^2\theta\cos 2\phi + \lambda_{\theta\phi}\sin 2\theta\cos\phi),
\end{equation*}
} 
\end{linenomath}
where $\theta$ and $\phi$ are the polar and azimuthal angles, respectively, of the 
positively charged muon with respect to the $z$ axis of a 
polarization frame, and $\lambda_\theta$, $\lambda_\phi$, and $\lambda_{\theta\phi}$ are the angular distribution parameters.
To estimate the effect of the polarization on the measurement of the $\PgUa\PgUa$ fiducial cross section, we choose to use the helicity frame, where the
polar axis coincides with the direction of the \PgUa momentum. Measurements performed by the CMS and LHCb Collaborations 
on single $\PgU$ production indicate compatibility of all
the angular distribution parameters with zero over a large phase space~\cite{Aaij:2017egv,Chatrchyan:2012woa}.
 However, the same may not be true for \PgUa pair production. To estimate the effect of polarization on the \PgUa pair 
production cross section, simulated events are reweighted to have the angular distributions corresponding to various 
$\lambda_\theta$ values, without changing the overall simulated yield. The same efficiency and acceptance 
corrections as in Eq.~(\ref{eq:corr}) are used to calculate \Ncorr for these different polarization scenarios. 
The variations in the measured \PgUa pair production cross section are given for different 
$\lambda_\theta$ coefficients in Table~\ref{tab:pola}. The effect of different polarizations can be substantial, changing the 
measured cross section by $-60$ to $+25\%$. 

\begin{table*}
\topcaption{Variation of the measured fiducial \PgUa pair production cross section for several $\lambda_\theta$ coefficient values.}
\label{tab:pola}
\centering
\begin{tabular}{lcccccccc}
\hline
$\lambda_\theta$ & $-1.0$ & $-0.5$ & $-0.3$ & $-0.1$ & $+0.1$ & $+0.3$ & $+0.5$ & $+1.0$ \\
$\Delta\sfid$ & $-60\%$ & $-22\%$ & $-12\%$ & $-3.7\%$ & $+3.4\%$ & $+9.4\%$ & $+14\%$ & $+25\%$  \\
\hline
\end{tabular}
\end{table*}

\subsection{Measurement of the DPS-to-inclusive fraction}
\label{sec:fdps}

The DPS and SPS mechanisms lead to different kinematic distributions for the $\PgUa\PgUa$ events. The DPS 
production is characterized by a larger separation in rapidity between the mesons, \deltay, as they are largely uncorrelated, and by
a larger invariant mass of the meson pairs, \mYY. The distributions of $\Delta\phi(\PgUa,\PgUa)$, $\Delta R(\PgUa,\PgUa)$, and $\pt(\PgUa\PgUa)$ also
differ for the SPS and DPS mechanisms, but they are very sensitive to the choice of model parameters in 
the simulation and are subject to large theoretical uncertainties~\cite{Cook:2140033}.
Measuring the $\PgUa\PgUa$ fiducial cross section in bins of \deltay or of \mYY can give
a measurement of the fraction of DPS events, \fdps, defined as:
\begin{linenomath} 
\begin{equation}
\fdps =\frac{\sfidDPS}{\sfidSPS+\sfidDPS},
\end{equation}
\end{linenomath} 
where \sfidDPS and \sfidSPS are, respectively, the DPS and SPS cross sections in the fiducial region.
We measure the fiducial cross section in five bins of \deltay 
and five bins of \mYY.
The signal and background models are the same as for the inclusive measurement, except that the width of the function 
describing the \PgUa invariant mass shape is allowed to float between its best-fit values for the inclusive selection and for the selection in the relevant exclusive bin. This allows for a potential degradation (improvement) of the muon momentum resolution at high (low) pseudorapidity to be taken into account, since
the muon pseudorapidity is correlated with both \deltay and \mYY. The systematic 
uncertainties are identical to those presented in Section~\ref{sec:systematics1}.

\begin{figure}[h!]
    \centering
        \includegraphics[width=0.47\textwidth]{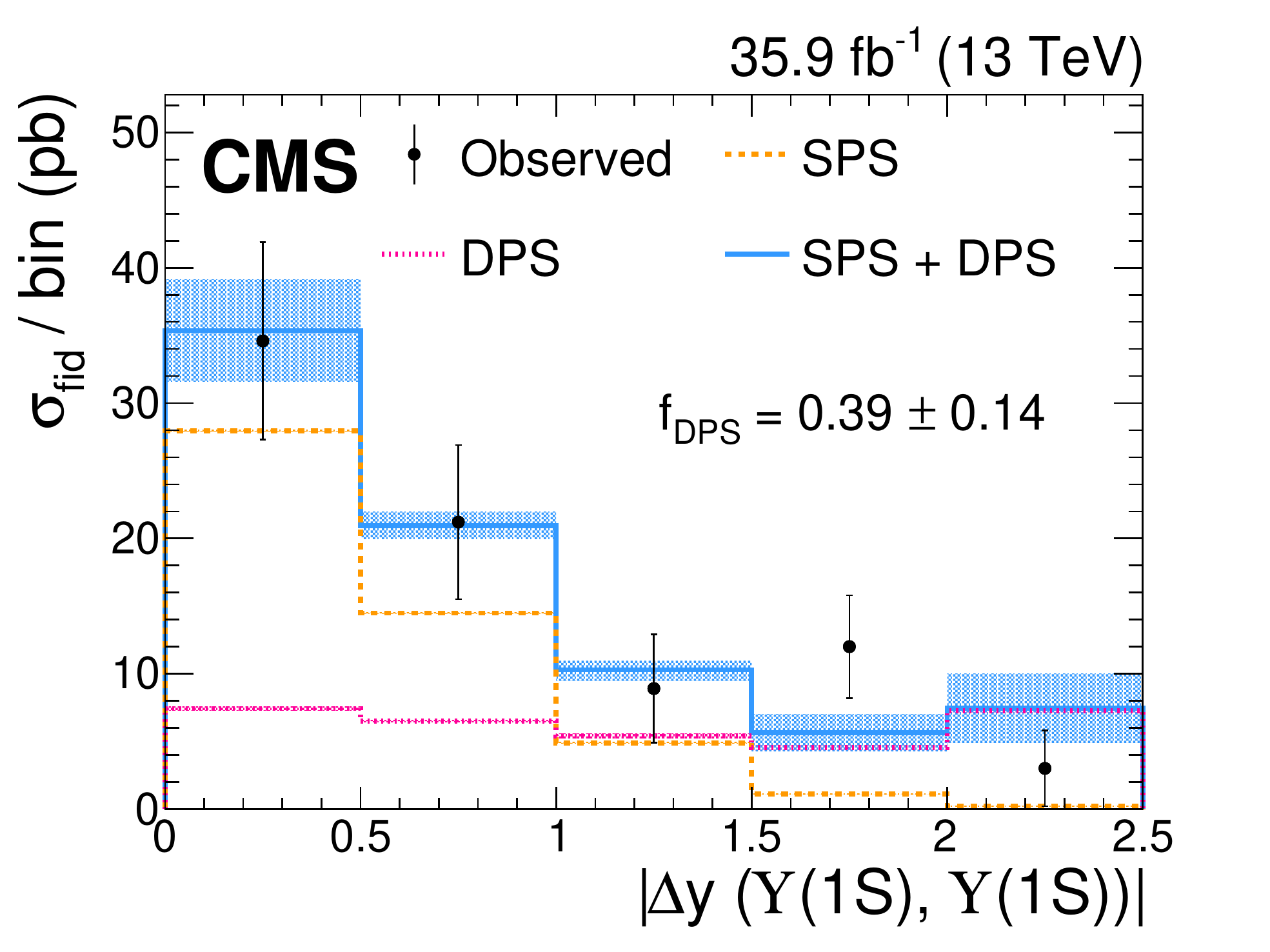}
        \includegraphics[width=0.47\textwidth]{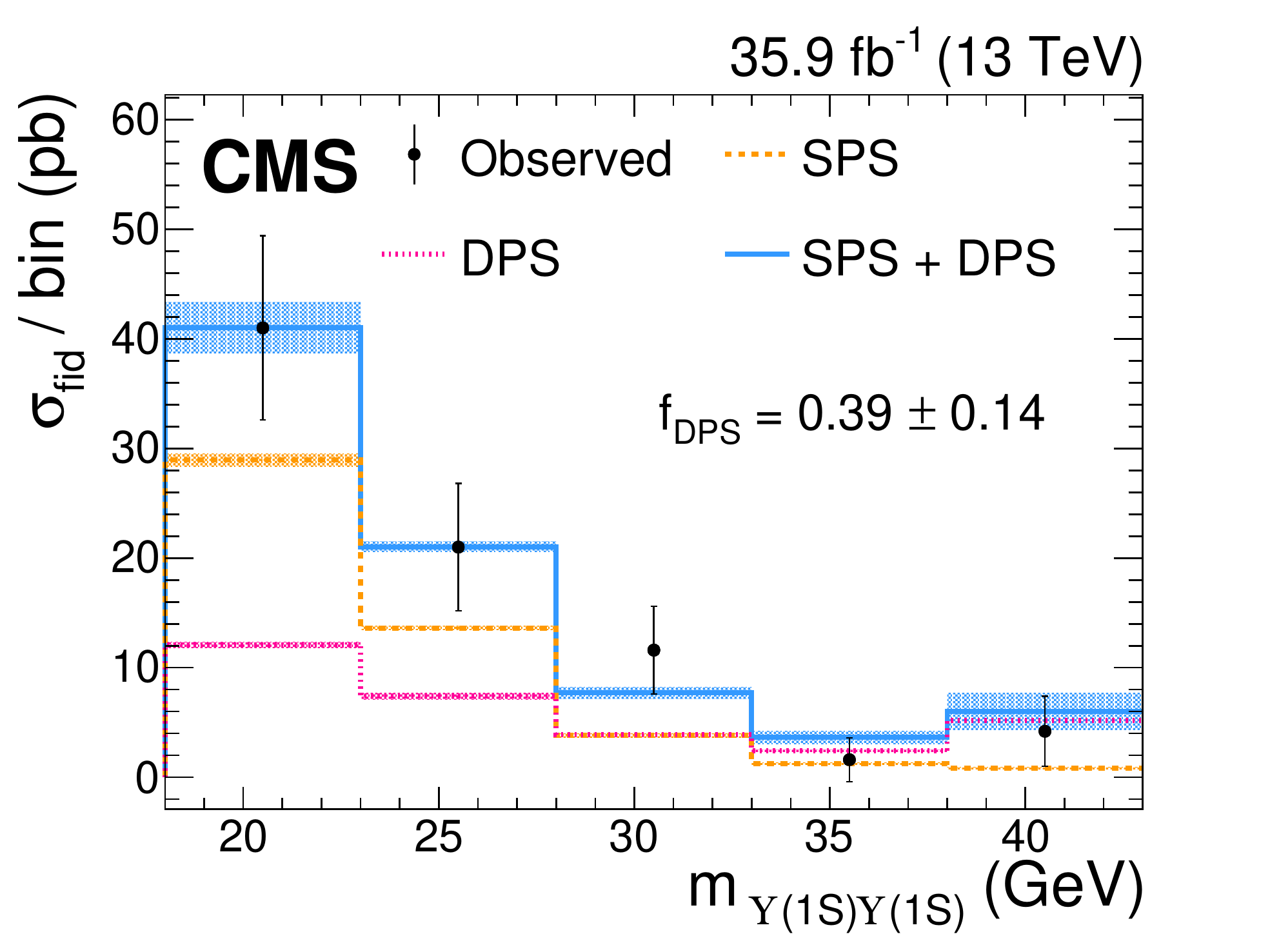}
    \caption{Measured fiducial cross section (black dots) in bins of \deltay (\cmsLeft) or \mYY (\cmsRight). The last bin includes the overflow. The SPS and DPS distributions predicted from simulation are overlaid using the \fdps value extracted from the fit to the \deltay distribution. The shaded areas around the SPS and DPS predictions indicate the theoretical uncertainties, which are often smaller than the thickness of the dashed lines. The shaded area around the total distribution corresponds to the uncertainty in the measurement of \fdps. The solid line shows the sum of the SPS and DPS contributions with the best-fit \fdps.}
    \label{fig:fdps}
\end{figure}

The extracted fiducial cross sections as a function of \deltay and \mYY are compared
to the expected distributions for SPS and DPS production, as obtained in the fiducial region using \PYTHIA for the DPS process, and from \textsc{HELAC-Onia} 
with the NLO* CSM predictions for the SPS process. The fraction \fdps is measured with a binned maximum-likelihood fit of
these two simulated distributions with floating normalizations to the measured fiducial cross sections in bins of \deltay and \mYY.
As determined from pseudo-experiments, the best precision is expected to be achieved using \deltay. 
Theoretical uncertainties coming from the choice of parton distribution functions and the factorization and renormalization scales are taken into account for 
both the SPS and DPS predicted distributions. 
The fraction \fdps is measured to be
$(39\pm 14)\%$ using \deltay as the discriminative distribution. This results includes both statistical 
and systematic uncertainties, where the former strongly dominates. 
The result using \mYY is compatible with this measurement, but with much lower precision:
$(27\pm22)\%$. The uncertainties are strongly dominated by the uncertainties in the measurements of the cross section in the 
 \deltay and \mYY bins, with theoretical uncertainties 
in the predicted SPS and DPS distributions playing a role at the percent level.  The measured differential fiducial cross sections are shown in Fig.~\ref{fig:fdps}, together with the
SPS and DPS predictions.

\section{Search for resonances}
\label{sec:search}

\subsection{Methodology}

We search for a narrow excess of events above an expected smooth four-muon invariant mass spectrum. Assuming that the resonant state decays into two muons and a \PgUa meson  that further decays to a pair of muons, the signal mass resolution 
can be improved by using a mass-difference observable~\cite{Sirunyan:2018iwt}: 
\begin{linenomath} 
\begin{equation}
\mtilde = \mmmmm - m_{\Pgm\Pgm} + m_{\PgUa},
\end{equation}
\end{linenomath} 
where $\mmmmm$ is the invariant mass of the four leptons, $m_{\Pgm\Pgm}$ the invariant mass associated with the \PgUa candidate, and $m_{\PgUa}$ the nominal mass 
of the \PgUa particle (9.46\GeV~\cite{PhysRevD.98.030001}). 
This estimated mass, denoted as \mtilde, has a resolution about 50\% better than the four-muon invariant mass \mmmmm for signal events. The \mmmmm and \mtilde distributions are similar for the combinatorial background.

The results are extracted by performing an unbinned maximum-likelihood fit to the \mtilde spectrum. The signal and background components are modeled by several functional forms in the fit, as described in the next paragraphs. 

The signal distributions are parameterized by the sum of two Gaussian functions with the same mean. 
The parameters are extracted for the four mass points available in simulation. The signal modeling needs to be interpolated 
for masses between 16.5 and 26\GeV and extrapolated to masses up to 27\GeV to search for narrow resonances with any mass 
between 16.5 and 27\GeV. This is done by fitting with polynomials the different parameters of the two Gaussian functions as a 
function of the generated resonance mass. 
The same procedure is repeated for every signal model. The full width at half maximum is about 0.2\GeV for a resonance mass of 18\GeV. 

The background is separated into two components: the $\PgUa\PgUa$ process, which was the signal in Section~\ref{sec:xs} and 
is characterized by a sharp rising edge in the \mtilde spectrum at twice the \PgUa meson mass, 
and the combinatorial background, which is described by a smooth function as explained below. 

\begin{figure}[hbpt]
\centering
        \includegraphics[width=0.49\textwidth]{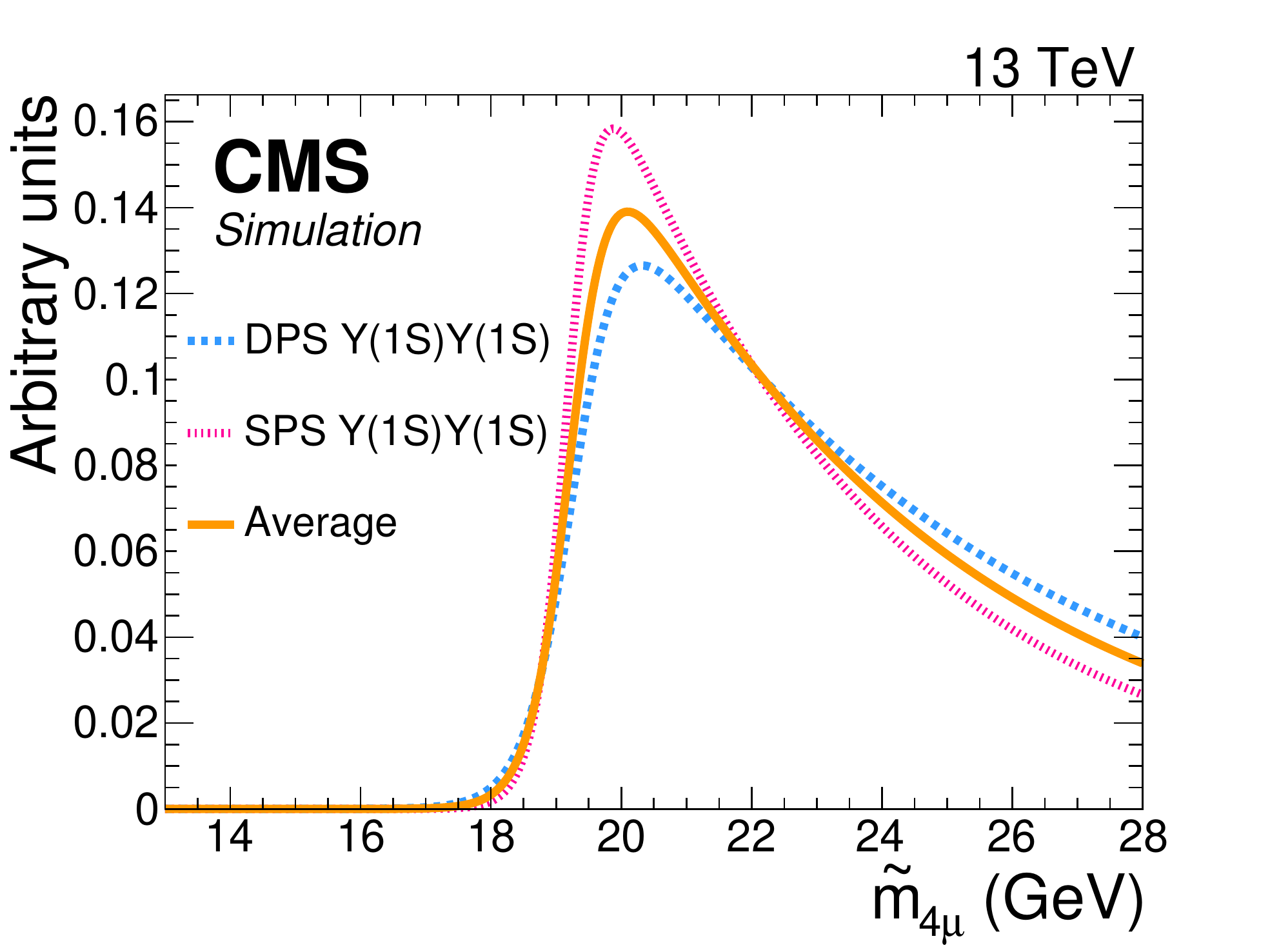}
    \caption{Distributions of $\mtilde$ for simulated $\PgUa\PgUa$ events. The dashed lines are the best-fit models for the SPS and DPS simulations. The solid line is the average of the SPS and DPS models, which is taken as the nominal model for the $\PgUa\PgUa$ background in the resonance search. } 
    \label{fig:backgroundYY}
\end{figure}

The \mtilde spectrum for the $\PgUa\PgUa$ process is obtained from simulation, 
and is modeled as the product of a sigmoid function and an exponential function with a negative exponent. The nominal model for the $\PgUa\PgUa$ background is taken as 
an average between the DPS and SPS templates, which is consistent with the measurement of the DPS fraction presented in Section~\ref{sec:results1}. 
Figure~\ref{fig:backgroundYY} 
shows the \mtilde models obtained from simulated DPS and SPS events, together with the average fit model.   
The number of $\PgUa\PgUa$ events 
in the signal region is extracted, as detailed in Section~\ref{sec:xs}, using the selection designed for 
the resonance search and without applying the acceptance and efficiency corrections from Eq.~(\ref{eq:corr}). In this case, 
only events with $13<\mtilde<28\GeV$ are retained and no rapidity criteria are applied for the 
reconstructed \PgUa candidates. The yield is measured to be $78\pm 13$ events. 
The requirement that the mass of a dimuon pair is compatible with the mass of a \PgUa meson within 
two standard deviations is enforced in the resonance search but is not applied to extract the yield because the 2D fit relies on the mass tails to estimate 
the combinatorial background. Since the efficiency of this criterion is 95\% in both the SPS and DPS $\PgUa\PgUa$ simulations, the $\PgUa\PgUa$ yield in the signal region is corrected to $74\pm13$. The normalization of the \PgUa pair production 
process and its uncertainty are extracted from the same data as in 
the signal region of the resonance search, but this does not lead to a significant overconstraint of the uncertainty in the 
maximum-likelihood fit of the \mtilde distribution because the latter can determine the \PgUa pair normalization 
only with poor precision.  

\begin{figure}[hbpt]
\centering
        \includegraphics[width=0.49\textwidth]{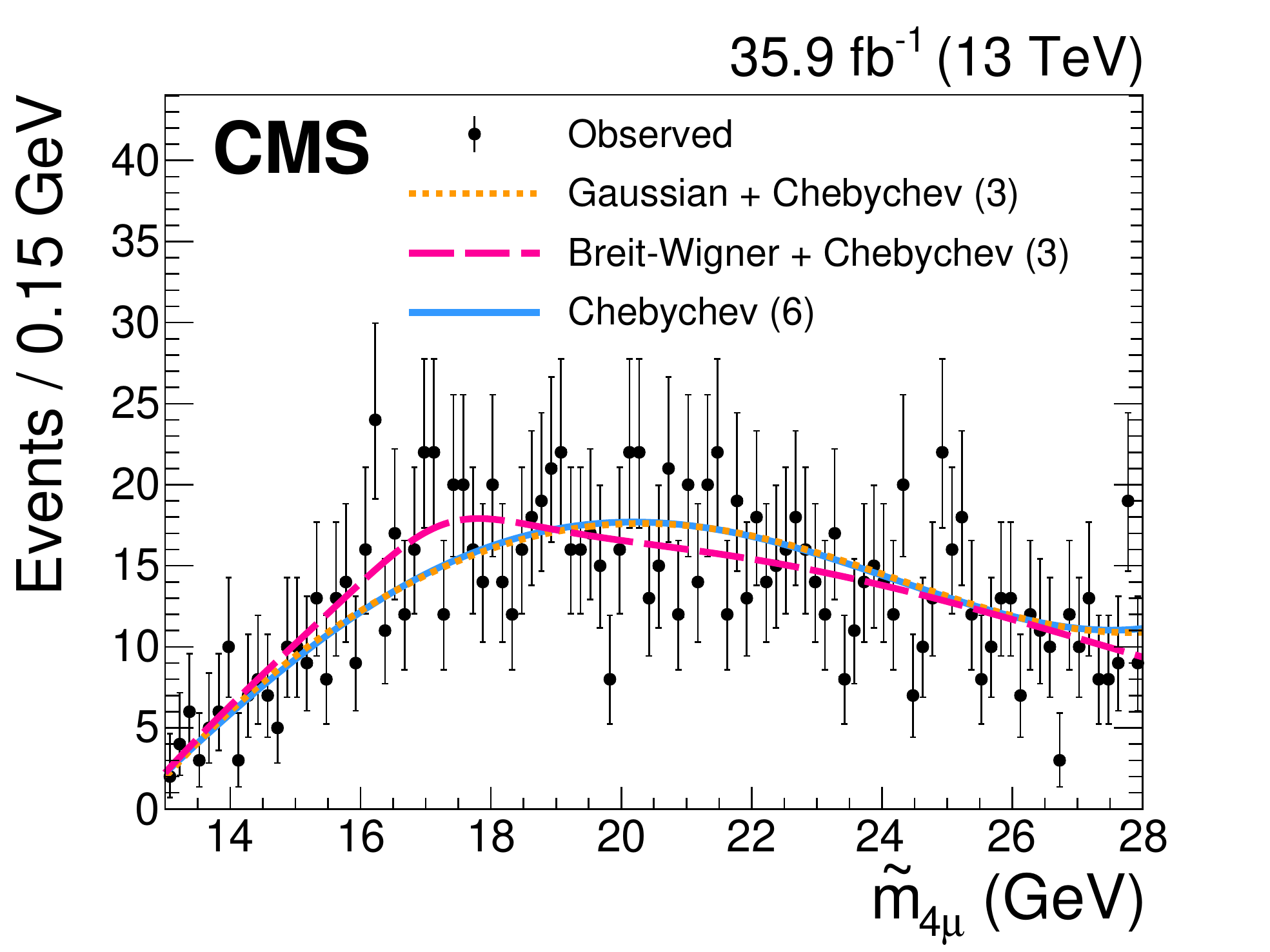}
    \caption{Distributions of \mtilde for the combinatorial background in a control region with the vertex fit $\chi^2$ probability of the four muons 
in the range $10^{-10}$--$10^{-3}$. The parameters obtained in this fit are not used as an input for the fit in the signal region. 
The functional forms for the combinatorial background shown by the lines are all considered as possible shapes 
for the background model in the likelihood fit. 
The order of the polynomials is indicated in parentheses in the legend. } 
    \label{fig:backgroundcmb}
\end{figure}

The \mtilde spectrum for the combinatorial background is obtained in the fit to the data in the signal region. Several generic functions are used to parameterize this smooth background:
\begin{itemize}
\item Chebychev polynomials of various orders;
\item the sum of a Gaussian function and a Chebychev polynomial;
\item the sum of a Breit--Wigner function and a Chebychev polynomial.
\end{itemize}
The widths of the Gaussian and Breit--Wigner functions are constrained to be above 2\GeV to avoid fitting narrow structures due to statistical fluctuations. 
We verify, using a control region where the 
vertex fit $\chi^2$ probability of the four muons is in the range $10^{-10}$--$10^{-3}$, that these three functional forms 
describe the smooth \mtilde spectrum of the combinatorial background with a good $\chi^2$ probability. Muons with a vertex probability in this 
range are likely to be associated with processes from the same primary vertex, but can originate from decays in flight or 
displaced secondary vertices. 
This control region is shown in Fig.~\ref{fig:backgroundcmb} for illustrative purposes. 
The parameters of the functions determined from the fit are not used in the signal region, where the parameters of the combinatorial background, as well 
as the choice of the functional form, are freely floating. 

\subsection{Systematic uncertainties}
\label{sec:systematics2}

The systematic uncertainties are to a large extent similar to those used in the measurement of the $\PgUa\PgUa$ cross section and introduced in 
Section~\ref{sec:systematics1}. In this section, only the differences are highlighted. 
They arise from slightly different selection criteria, a different choice of observable, the treatment of the $\PgUa\PgUa$ process 
as a background, and the introduction of a new signal process. 

The uncertainty per muon in the muon identification and tracking is increased from 0.5\% to 1\% because poorly reconstructed muons with $\pt<3.5\GeV$ in the barrel are 
included in the resonance search to increase the signal acceptance for light resonances. 
In addition, in the resonance search, the signal is affected by a 1\% yield 
uncertainty related to the requirement that the \PgUa candidate has an invariant mass compatible with the nominal \PgUa meson mass within two 
standard deviations. This uncertainty is determined by comparing the dimuon invariant mass resolution distributions in $\PgUa\PgUa$ simulated events and in \PgUa events in data.
The modeling of the signal process with a resonance mass other than those for which simulated samples were generated leads to a 2\% uncertainty
in the signal normalization for the resonance search.

The discrete profiling method~\cite{Dauncey:2014xga} is used to model the combinatorial background. 
This allows the choice of the fit functions among those provided to be considered as a discrete nuisance parameter. 
The parameters of these fit functions are freely floating. 

The normalization of the $\PgUa\PgUa$ background in the resonance search is extracted from the 2D unbinned fit to the 
invariant mass of the dimuon pairs in the \PgUa mass region. The uncertainty in the yield obtained from the fit is considered 
as a log-normal uncertainty in the fit to the \mtilde distribution. The \mtilde distribution of the $\PgUa\PgUa$ background is allowed to float between the predictions 
for the SPS and DPS simulations. 

Uncertainties in the \mtilde distribution of the resonant signal 
take into account the limited size of the simulated samples, and the limited precision of the description of the signal for masses not available in simulations. The uncertainty in the mean mass of the signal is 0.2\%, 
corresponding to the uncertainty in the muon momentum scale. The other parameters describing the shape of 
the signal have an uncertainty between 5 and 15\%, which leads to a combined impact on the final upper limits of less than 2\%. 

The uncertainty in the \PgUa dimuon branching fraction is not considered, since the limits are set on the product of the resonance production cross section and its branching fraction to four muons via an intermediate \PgUa resonance.

\subsection{Results}
\label{sec:results2}

\begin{figure}[h!]
    \centering
        \includegraphics[width=0.49\textwidth]{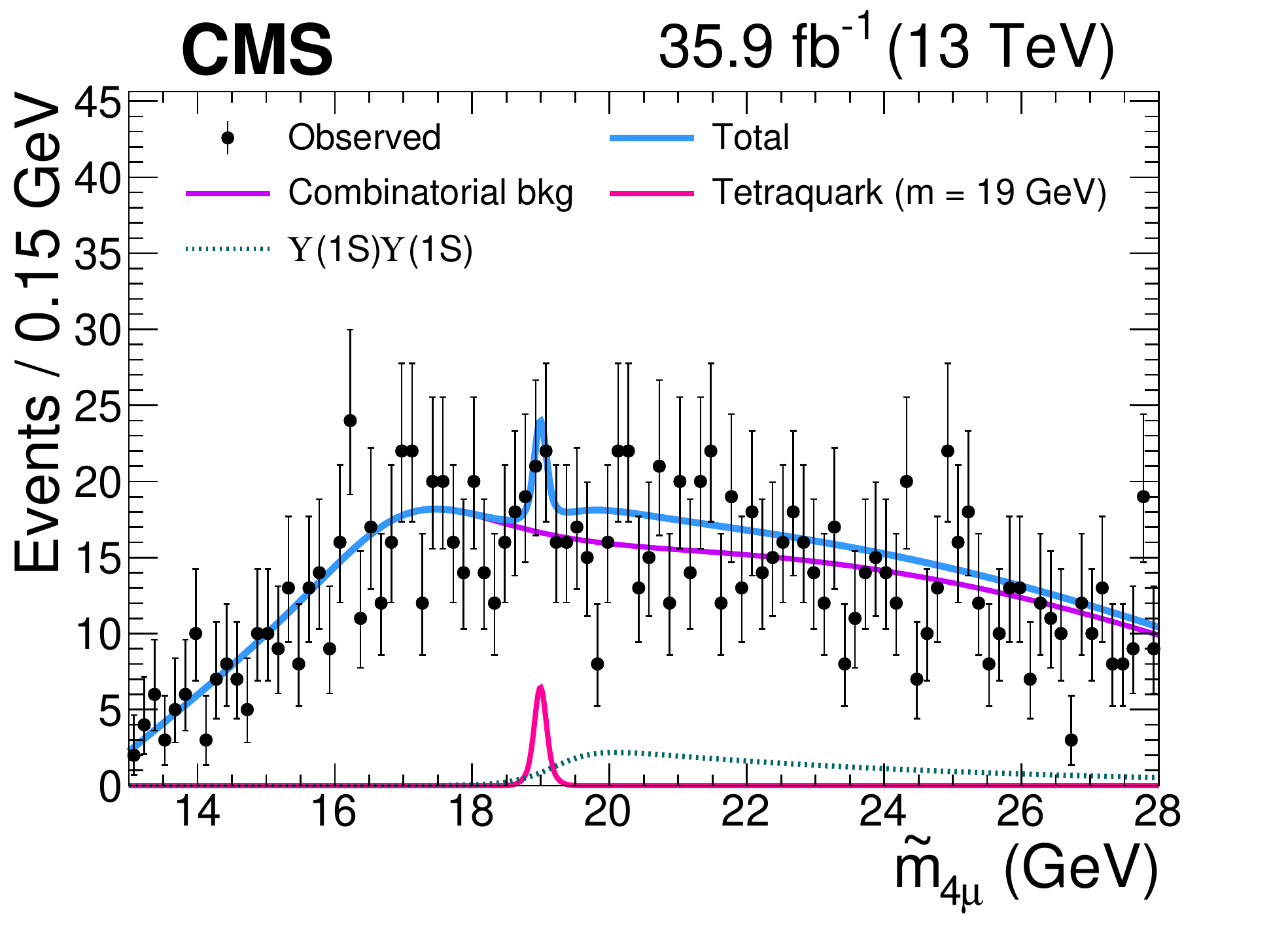}
\caption{The \mtilde distribution from data and the results of the fit in the resonance search. An example signal is shown for the tetraquark model with a mass of 19\GeV, which has a significance of about one standard deviation. }
    \label{fig:scan}  
\end{figure}

The binned \mtilde distribution in the signal region of the resonance search is shown in Fig.~\ref{fig:scan}. The background and example signal components are 
shown using their best-fit shapes and normalizations.  
Using the number of $\PgUa\PgUa$ events observed in data as a reference, a resonance with a mass around 19\GeV and 
having a similar production cross section and branching fraction to four muons as the $\PgUa\PgUa$ production, 
would produce about 100 events in our sample, given the similarity between the kinematic distributions of both processes.
No significant narrow excess of events is
observed above the background expectation. 
The largest excess is observed for a resonance mass of 25.1\GeV, and has a local significance of 2.4 standard deviations for the scalar signal hypothesis.

\begin{figure*}[h!]
    \centering
        \includegraphics[width=0.47\textwidth]{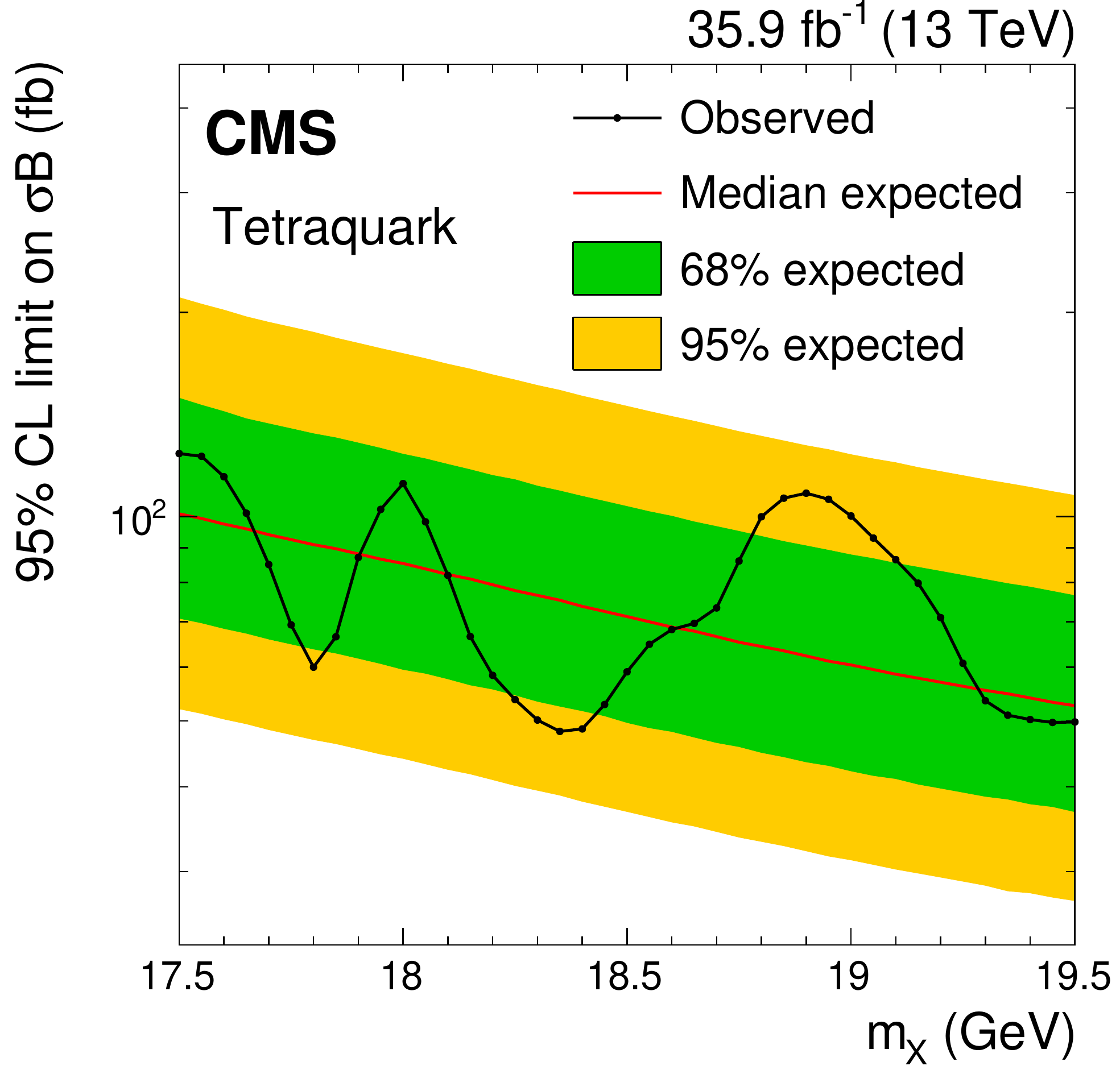}
        \includegraphics[width=0.47\textwidth]{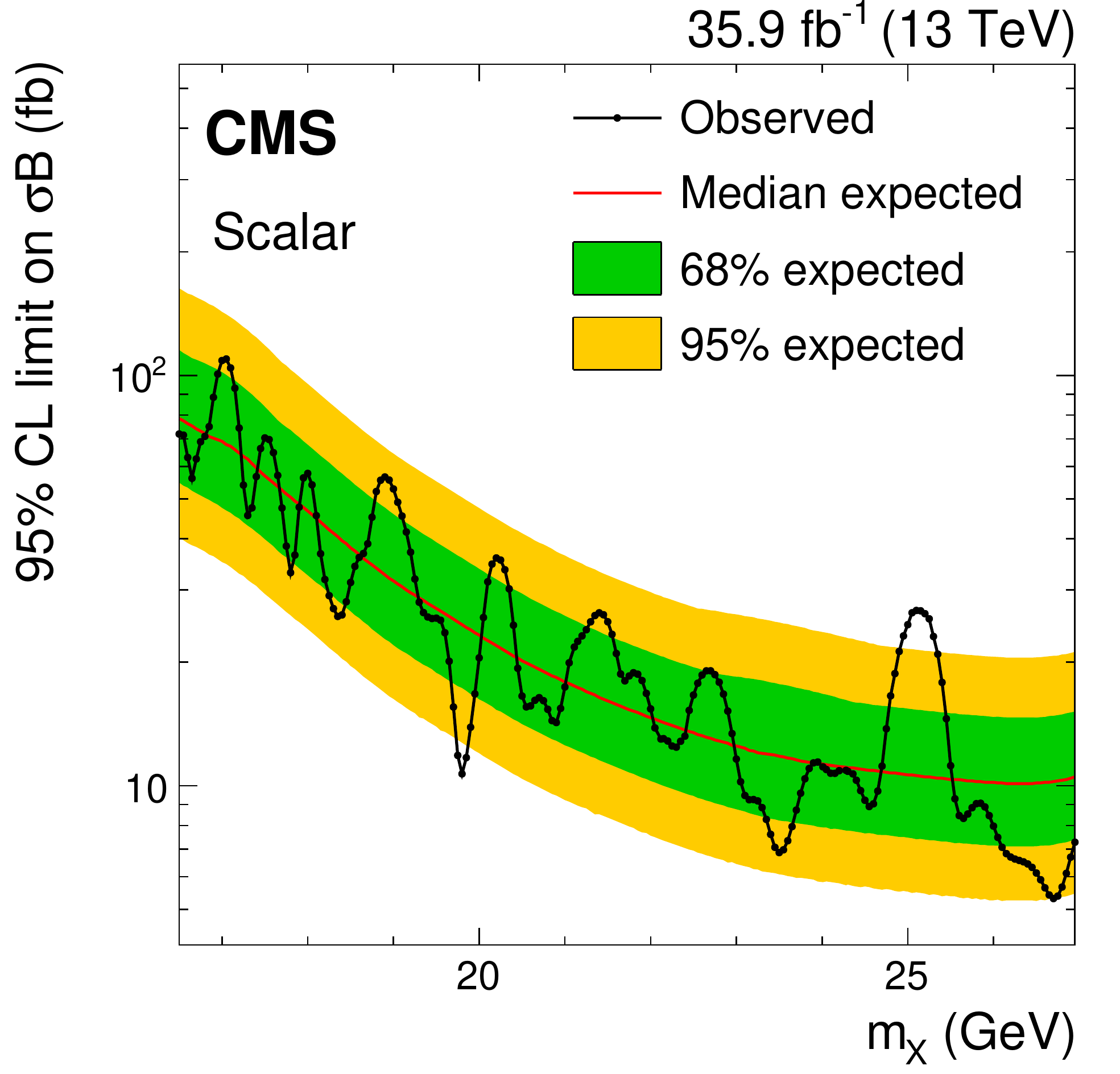}\\
        \includegraphics[width=0.47\textwidth]{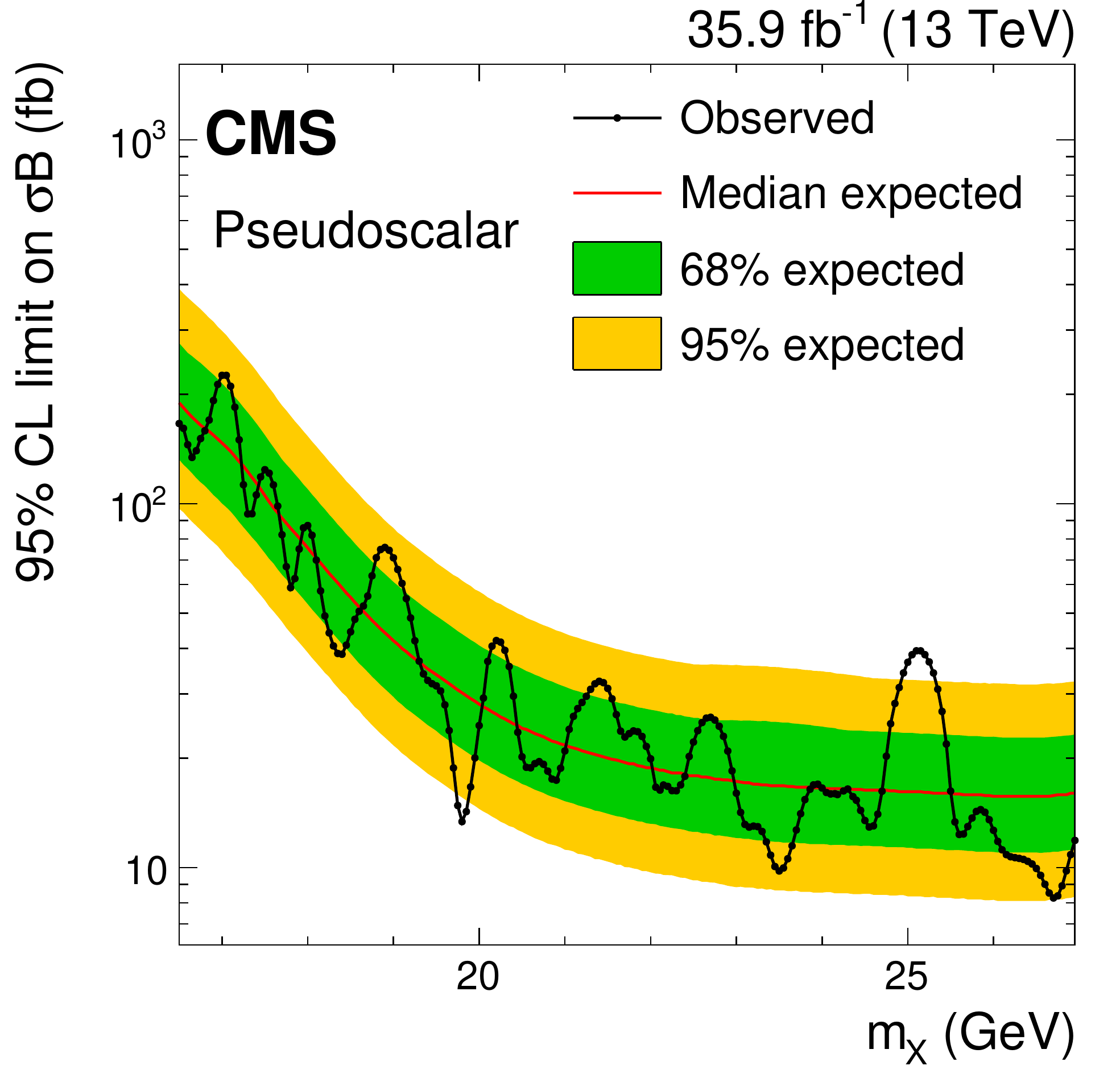}
        \includegraphics[width=0.47\textwidth]{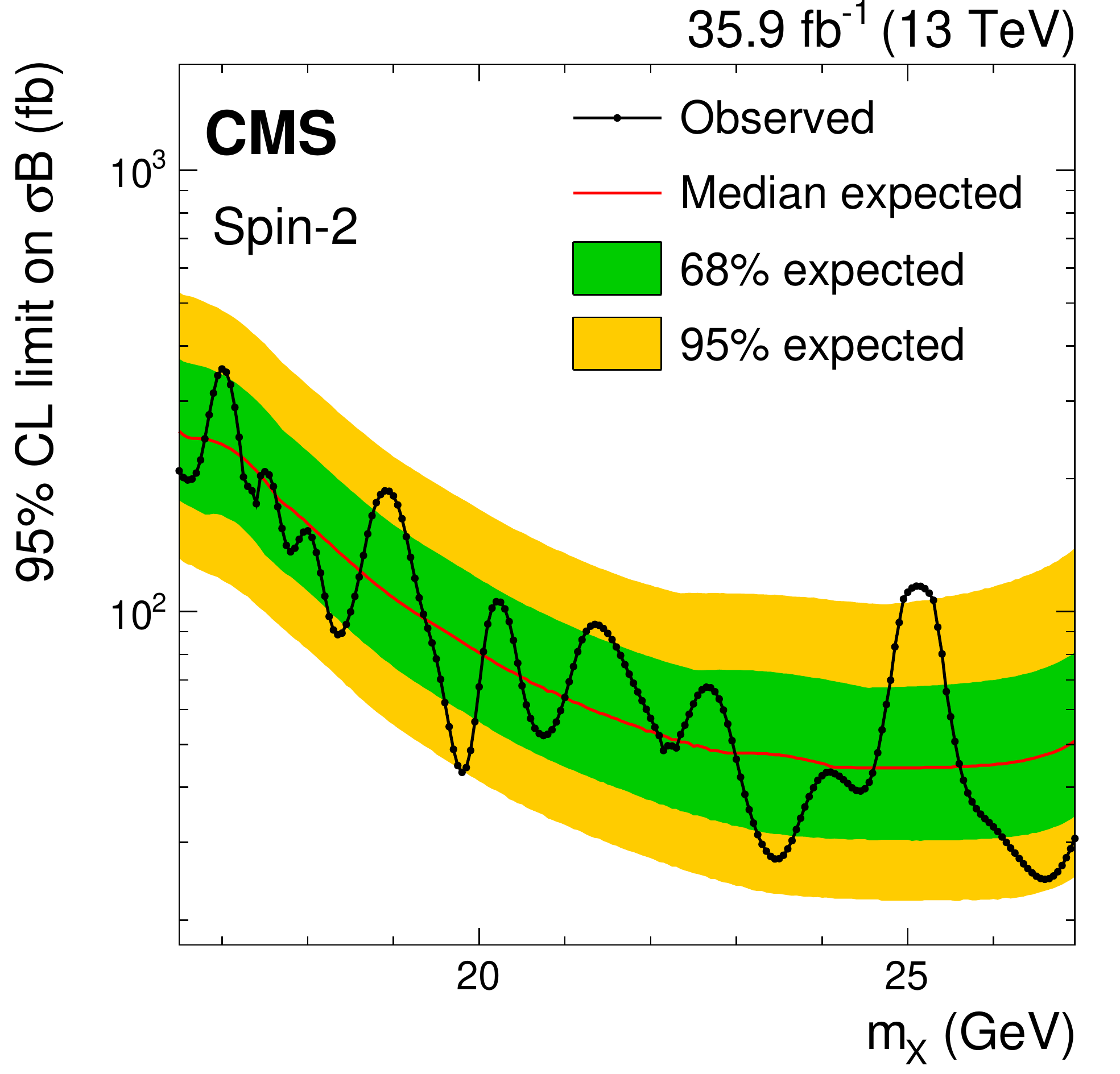}\\
    \caption{Upper limits at 95\% \CL on the product of the cross section and branching fraction for a tetraquark (upper left), scalar (upper right), 
pseudoscalar (lower left), and spin-2 (lower right) states. The symbol $\sigma$ denotes the production cross section of the resonance, and the symbol $\mathcal{B}$ denotes the product of the branching fraction for the decay of the resonance to a $\PgUa$ meson and two muons, and the $\PgUa$ meson dimuon branching fraction. The line with the points on it shows the observed upper limits and the thin red line is the median of the expected upper limits. The inner (green) band and the outer (yellow) band indicate the regions containing 68 and 95\%, respectively, of the distribution of limits expected under the background-only hypothesis. }
    \label{fig:limits}
\end{figure*}

Upper limits on the product of the production cross section of a resonance and the branching fraction to a final state of four muons via 
an intermediate \PgUa resonance are set at 95\% confidence level (CL) using the modified frequentist
 construction \CLs in the asymptotic approximation~\cite{LHC-HCG-Report,Chatrchyan:2012tx,Junk,Read:2002hq,Cowan:2010js}, 
separately for each signal model. The upper limits are extracted using unbinned distributions. The cross section is defined in the entire phase space without fiducial requirements, 
and the branching fraction used is the product of the branching fraction of the resonant state to a \PgUa meson and two muons, and the  
branching fraction of the \PgUa meson to two muons. 
Masses between 17.5 and 19.5\GeV are probed in the context of the tetraquark search, using the bottomonium model, 
whereas the limits in the extended mass range 16.5--27\GeV are set for the generic search, using the \textsc{JHUGen} models.
The corresponding upper limits are given in Fig.~\ref{fig:limits}. They range between 5 and 380\unit{fb}, depending on the mass and 
 signal model. The patterns in the limits are broader for the spin-2 signal than for the scalar and pseudoscalar 
states because the signal is characterized by softer and more forward muons, leading to a worse $\mtilde$ resolution.

\section{Summary}

The cross section for \PgUa pair production is measured in the fiducial 
region where both \PgUa mesons have an absolute rapidity below 2.0. 
The measurement is performed using proton-proton collision data collected at a center-of-mass energy of 13\TeV by the CMS detector in 2016 and corresponding to an integrated luminosity of 35.9\fbinv. 
Assuming that the \PgUa mesons are produced unpolarized, the fiducial \PgUa pair production cross section 
is determined to be $79 \pm 11\stat \pm 6\syst \pm 3(\mathcal{B})$\unit{pb}, where 
the last uncertainty comes from the 
uncertainty in the \PgUa dimuon branching fraction. The result can change if the \PgUa  mesons are 
produced with a nonzero polarization. Changing the polarization 
coefficient $\lambda_\theta$ from $-1$ to $+1$, the resulting \PgUa pair production cross section 
measurement varies by $-60$ to $+25\%$.

The contribution of double-parton scattering to the total inclusive \PgUa pair production cross section is determined 
for the first time. It is measured to be ($39\pm14$)\% in the same fiducial region as described above, 
where the uncertainty includes both statistical and systematic components, with the 
statistical uncertainty dominating. 

The results of a search are also presented for a light narrow resonance, such as a tetraquark or a bound state beyond-the-standard model, decaying to a \PgUa and a pair of opposite-sign muons.
No excess of events compatible with a signal is observed in the four-muon invariant mass spectrum. Upper limits at 95\% confidence level on the product of the signal cross section and branching fraction to four muons via an intermediate \PgUa resonance are set for different signal models, expanding the kinematic and mass coverage of previous searches.   

\begin{acknowledgments}
    We congratulate our colleagues in the CERN accelerator departments for the excellent performance of the LHC and thank the technical and administrative staffs at CERN and at other CMS institutes for their contributions to the success of the CMS effort. In addition, we gratefully acknowledge the computing centers and personnel of the Worldwide LHC Computing Grid for delivering so effectively the computing infrastructure essential to our analyses. Finally, we acknowledge the enduring support for the construction and operation of the LHC and the CMS detector provided by the following funding agencies: BMBWF and FWF (Austria); FNRS and FWO (Belgium); CNPq, CAPES, FAPERJ, FAPERGS, and FAPESP (Brazil); MES (Bulgaria); CERN; CAS, MoST, and NSFC (China); COLCIENCIAS (Colombia); MSES and CSF (Croatia); RPF (Cyprus); SENESCYT (Ecuador); MoER, ERC IUT, PUT and ERDF (Estonia); Academy of Finland, MEC, and HIP (Finland); CEA and CNRS/IN2P3 (France); BMBF, DFG, and HGF (Germany); GSRT (Greece); NKFIA (Hungary); DAE and DST (India); IPM (Iran); SFI (Ireland); INFN (Italy); MSIP and NRF (Republic of Korea); MES (Latvia); LAS (Lithuania); MOE and UM (Malaysia); BUAP, CINVESTAV, CONACYT, LNS, SEP, and UASLP-FAI (Mexico); MOS (Montenegro); MBIE (New Zealand); PAEC (Pakistan); MSHE and NSC (Poland); FCT (Portugal); JINR (Dubna); MON, RosAtom, RAS, RFBR, and NRC KI (Russia); MESTD (Serbia); SEIDI, CPAN, PCTI, and FEDER (Spain); MOSTR (Sri Lanka); Swiss Funding Agencies (Switzerland); MST (Taipei); ThEPCenter, IPST, STAR, and NSTDA (Thailand); TUBITAK and TAEK (Turkey); NASU (Ukraine); STFC (United Kingdom); DOE and NSF (USA). 

    \hyphenation{Rachada-pisek} Individuals have received support from the Marie-Curie program and the European Research Council and Horizon 2020 Grant, contract Nos.\ 675440, 752730, and 765710 (European Union); the Leventis Foundation; the A.P.\ Sloan Foundation; the Alexander von Humboldt Foundation; the Belgian Federal Science Policy Office; the Fonds pour la Formation \`a la Recherche dans l'Industrie et dans l'Agriculture (FRIA-Belgium); the Agentschap voor Innovatie door Wetenschap en Technologie (IWT-Belgium); the F.R.S.-FNRS and FWO (Belgium) under the ``Excellence of Science -- EOS" -- be.h project n.\ 30820817; the Beijing Municipal Science \& Technology Commission, No. Z191100007219010; the Ministry of Education, Youth and Sports (MEYS) of the Czech Republic; the Deutsche Forschungsgemeinschaft (DFG) under Germany’s Excellence Strategy -- EXC 2121 ``Quantum Universe" -- 390833306; the Lend\"ulet (``Momentum") Program and the J\'anos Bolyai Research Scholarship of the Hungarian Academy of Sciences, the New National Excellence Program \'UNKP, the NKFIA research grants 123842, 123959, 124845, 124850, 125105, 128713, 128786, and 129058 (Hungary); the Council of Science and Industrial Research, India; the HOMING PLUS program of the Foundation for Polish Science, cofinanced from European Union, Regional Development Fund, the Mobility Plus program of the Ministry of Science and Higher Education, the National Science Center (Poland), contracts Harmonia 2014/14/M/ST2/00428, Opus 2014/13/B/ST2/02543, 2014/15/B/ST2/03998, and 2015/19/B/ST2/02861, Sonata-bis 2012/07/E/ST2/01406; the National Priorities Research Program by Qatar National Research Fund; the Ministry of Science and Education, grant no. 14.W03.31.0026 (Russia); the Programa Estatal de Fomento de la Investigaci{\'o}n Cient{\'i}fica y T{\'e}cnica de Excelencia Mar\'{\i}a de Maeztu, grant MDM-2015-0509 and the Programa Severo Ochoa del Principado de Asturias; the Thalis and Aristeia programs cofinanced by EU-ESF and the Greek NSRF; the Rachadapisek Sompot Fund for Postdoctoral Fellowship, Chulalongkorn University and the Chulalongkorn Academic into Its 2nd Century Project Advancement Project (Thailand); the Kavli Foundation; the Nvidia Corporation; the SuperMicro Corporation; the Welch Foundation, contract C-1845; and the Weston Havens Foundation (USA). 
\end{acknowledgments}

\bibliography{auto_generated}   

\cleardoublepage \appendix\section{The CMS Collaboration \label{app:collab}}\begin{sloppypar}\hyphenpenalty=5000\widowpenalty=500\clubpenalty=5000\vskip\cmsinstskip
\textbf{Yerevan Physics Institute, Yerevan, Armenia}\\*[0pt]
A.M.~Sirunyan$^{\textrm{\dag}}$, A.~Tumasyan
\vskip\cmsinstskip
\textbf{Institut f\"{u}r Hochenergiephysik, Wien, Austria}\\*[0pt]
W.~Adam, F.~Ambrogi, T.~Bergauer, M.~Dragicevic, J.~Er\"{o}, A.~Escalante~Del~Valle, M.~Flechl, R.~Fr\"{u}hwirth\cmsAuthorMark{1}, M.~Jeitler\cmsAuthorMark{1}, N.~Krammer, I.~Kr\"{a}tschmer, D.~Liko, T.~Madlener, I.~Mikulec, N.~Rad, J.~Schieck\cmsAuthorMark{1}, R.~Sch\"{o}fbeck, M.~Spanring, W.~Waltenberger, C.-E.~Wulz\cmsAuthorMark{1}, M.~Zarucki
% ...
\vskip\cmsinstskip
\dag: Deceased\\
1:  Also at Vienna University of Technology, Vienna, Austria\\
% ...
\end{sloppypar}
\end{document}